\documentclass[a4paper,twocolumn,11pt,unpublished]{quantumarticle}
\pdfoutput=1
\usepackage[utf8]{inputenc}
\usepackage[english]{babel}
\usepackage[T1]{fontenc}
\usepackage{amsmath}
\usepackage{hyperref}
\usepackage{braket}

\usepackage{tikz}
\usepackage{lipsum}

\begin{document}

\title{A Jordan-Wigner gadget that reduces T count by more than 6x for quantum chemistry applications}

\author{Sam Pallister}
\affiliation{PsiQuantum, Palo Alto}
\orcid{0000-0003-1206-6296}
\email{spallister@psiquantum.com}
\maketitle

\begin{abstract}
  Quantum computers have the potential to be a profoundly transformative technology, particularly in the context of quantum chemistry. However, running a chemistry application that is demonstrably useful currently requires a prohibitive number of logical operations. For example, the canonical estimate of the number of operations required to simulate the molecule FeMoco, the key component in biological nitrogen fixation, requires around $10^{15}$ logical gates~\cite{Reiher2017}. A quantum computer that is capable of applying logical operations at 1~Mhz rates would require more than 30 years to complete such a calculation. It is imperative to reduce this prohibitive runtime, by better understanding and optimising quantum algorithms, if the technology is to have commercial utility. The purpose of this paper is to introduce such an optimisation. The gadget that we introduce below affords a 6x improvement in runtime for Trotterized quantum chemistry employing the Jordan-Wigner transformation, without altering the required number of qubits.
\end{abstract}

\vspace{0.3cm}
Upon completion of this manuscript, we became aware of the independent discovery of this result in~\cite{Wang2020}.
\vspace{0.3cm}

Quantum algorithms for quantum chemistry come in a variety of flavours, depending on the available hardware and problem instance. Algorithms such as the variation quantum eigensolver (VQE) are designed to run on near-term hardware, and extract meaningful chemical information by repeated measurement of expectation values given some circuit ansatz. Algorithms for fully fault-tolerant hardware, on the other hand, often rely on phase estimation to extract quantities of interest from the molecular Hamiltonian. Our focus will be on these algorithms.

Much like~\cite{Reiher2017}, we will focus on a particular subfamily of these quantum chemistry algorithms: those based on Trotter-Suzuki formulae (also known as ``product formulae'')~\cite{Suzuki1991}. Other families exist, such as those based on \emph{Qubitization}~\cite{Low2019}. While it may be the case that similar tricks to the one below have applicability outside of the Trotterization approach, we will not focus on them here. While much recent effort has been spent on the development and optimization of quantum simulation algorithms based on the qubitization framework~\cite{Babbush2018, Berry2019}, it is still unclear whether such an approach affords a universal improvement over Trotterization approaches based on low-order product formulae~\cite{Kivlichan2019, Childs2019}.

Additionally, we will assume the map that converts the formulation of the electronic structure problem in terms of electrons to that in terms of qubits is given by the \emph{Jordan-Wigner transformation}. Unlike the debate over Trotterization and Qubitization, the Jordan-Wigner transformation is used almost universally in quantum algorithms for this problem.

In Section~\ref{sec:jw}, we will review the Jordan-Wigner transformation in the context of electronic structure problems, as well as estimate the total cost of its implementation on a fault-tolerant quantum computer. In Section~\ref{sec:gadget}, we will introduce a compilation gadget that reduces the cost of this implementation by a significant constant factor.

\section{The Jordan-Wigner transformation}
\label{sec:jw}

The input to the Jordan-Wigner transformation is a fermionic Hamiltonian of the following form:
\begin{equation}
    H = \sum_{p,q=1}^M h_{pq} a^\dagger_p a_q + \sum_{p,q,r,s=1}^M h_{pqrs} a^\dagger_p a^\dagger_q a_r a_s,
\end{equation}
where $h_{pq}$ and $h_{pqrs}$ are real scalars and $a^\dagger_p$, $a_p$ are fermionic creation and annihilation operators for an electron in orbital $p$. Intuitively, the term $a^\dagger_p a_q$ corresponds to an electron moving from an orbital indexed by $q$ to orbital indexed by $p$; the relative propensity for this to happen is governed by the coefficient $h_{pq}$.

The coefficients $h_{pq}$ and $h_{pqrs}$ are derived from evaluating integrals given the molecular geometry and orbital basis. It is often the case that there is sufficient structure in the coefficients $h_{pq}$ and $h_{pqrs}$ to cause particular subsets of terms in the Hamiltonian to vanish. However, for large Hamiltonians, it is still the case that almost all of the terms will be of the form $a^\dagger_p a^\dagger_q a_r a_s, \; p \neq q \neq r \neq s$. Hence, we will focus on optimising the compilation of these terms in the main body of this paper, and defer discussion of the gadgetised form of the other terms to Appendix~\ref{sec:other_terms}. We also assume the ordering $p < q < r < s$, without loss of generality. 

Our goal is to extract the cost of implementing a term of this type, as part of a Trotterized quantum chemistry algorithm, on a fault-tolerant quantum computer that encodes its logical qubits using the surface code. In this regime, the fundamental quantity of interest is the number of non-Clifford operations that need to be applied to instantiate a given circuit; in particular, we count the temporal cost of the circuit in terms of T gates.

Each term in the fermionic Hamiltonian is mapped using Jordan-Wigner to a handful of terms acting on a qubit register of size $M$. The Jordan-Wigner transformation applies the following map:
\begin{align}\label{eq:jw}
    a_p \rightarrow \frac{1}{2} X_p \otimes \bigotimes_{t < p} Z_t + \frac{i}{2} Y_p \otimes \bigotimes_{t < p} Z_t; \\
    a^\dagger_p \rightarrow \frac{1}{2} X_p \otimes \bigotimes_{t < p} Z_t - \frac{i}{2} Y_p \otimes \bigotimes_{t < p} Z_t;
\end{align}
where $X_i$, $Y_i$ and $Z_i$ are Pauli matrices acting on qubit $i$.

As each creation and annihilation operator is mapped to a sum of two terms, the Hamiltonian term $h_{pqrs} a^\dagger_p a^\dagger_q a_r a_s$ therefore gets mapped to a sum of 16 terms. However, the electronic structure Hamiltonian is Hermitian and has real coefficients, and so this term must have a conjugate companion $(h_{pqrs} a^\dagger_p a^\dagger_q a_r a_s)^\dagger = h_{pqrs} a^\dagger_s a^\dagger_r a_q a_p$. Naively, one might think that the sum $\tilde{H} = h_{pqrs} (a^\dagger_p a^\dagger_q a_r a_s + a^\dagger_s a^\dagger_r a_q a_p)$ is mapped to 32 terms; however, it can be checked (either by using Eq.~\ref{eq:jw} or by commuting through annihilation operators) that all but 8 of these terms cancel, leaving the following:
\begin{widetext}
    \begin{alignat}{4}
        \nonumber \tilde{H} \rightarrow \tilde{H}_{JW} = h_{pqrs} \Big(- X_p &\otimes X_q &&\otimes X_r &&\otimes X_s &&\otimes \bigotimes_{\substack{p<t<q \\ r<t<s}} Z_t \otimes \bigotimes_{\text{elsewhere}} I \\
        \nonumber + X_p &\otimes X_q &&\otimes Y_r &&\otimes Y_s &&\otimes \bigotimes_{\substack{p<t<q \\ r<t<s}} Z_t \otimes \bigotimes_{\text{elsewhere}} I \\
        \nonumber - X_p &\otimes Y_q &&\otimes X_r &&\otimes Y_s &&\otimes \bigotimes_{\substack{p<t<q \\ r<t<s}} Z_t \otimes \bigotimes_{\text{elsewhere}} I \\
        \nonumber - X_p &\otimes Y_q &&\otimes Y_r &&\otimes X_s &&\otimes \bigotimes_{\substack{p<t<q \\ r<t<s}} Z_t \otimes \bigotimes_{\text{elsewhere}} I \\
        \nonumber - Y_p &\otimes X_q &&\otimes X_r &&\otimes Y_s &&\otimes \bigotimes_{\substack{p<t<q \\ r<t<s}} Z_t \otimes \bigotimes_{\text{elsewhere}} I \\
        \nonumber - Y_p &\otimes X_q &&\otimes Y_r &&\otimes X_s &&\otimes \bigotimes_{\substack{p<t<q \\ r<t<s}} Z_t \otimes \bigotimes_{\text{elsewhere}} I \\
        \nonumber + Y_p &\otimes Y_q &&\otimes X_r &&\otimes X_s &&\otimes \bigotimes_{\substack{p<t<q \\ r<t<s}} Z_t \otimes \bigotimes_{\text{elsewhere}} I \\
        - Y_p &\otimes Y_q &&\otimes Y_r &&\otimes Y_s &&\otimes \bigotimes_{\substack{p<t<q \\ r<t<s}} Z_t \otimes \bigotimes_{\text{elsewhere}} I \Big).
    \end{alignat}
\end{widetext}

Trotterization requires the application of unitaries corresponding to Hamiltonians of this form, exponentiated; i.e. we wish to implement a unitary of the form $\exp\{i\gamma \tilde{H}_{JW}\}$. However, if we use the Trotterized evolution as a component of phase estimation, we are actually interested in applying a \emph{controlled} version of this unitary, conditioned on the state of an ancilla upon which we wish to kick back a phase.

Exponentiating the Hamiltonian $\tilde{H}_{JW}$ can be achieved by exponentiating each term in turn, as they mutually commute. Each exponentiated term is a \emph{Pauli Product Rotation} (PPR); a unitary of the form $\exp\{i\alpha P\}$ for some angle $\alpha$ and string of Paulis $P$~\cite{Litinski2019}. Enforcing that each of these PPRs is controlled on the state of an ancilla incurs a 2x penalty in PPR count, as each controlled PPR can be constructed by two regular PPRs~\cite{Litinski2019}. Hence, the naive total cost in PPRs of applying $\exp\{i\gamma \tilde{H}_{JW}\}$ is 16. 

Each PPR has a cost in terms of T gates equivalent to a single qubit $R_z$, with arbitrary angle. Each of these rotations must be synthesised from T gates using an appropriate synthesis subroutine~\cite{Bocharov2015, Ross2016}. For example, the T count of synthesising a single-qubit rotation up to accuracy $\epsilon$ in the operator norm using the subroutine in~\cite{Ross2016} is approximately
\begin{equation}
    n_{T}^{synth}  = 3\log\left(\frac{1}{\epsilon}\right) + O\left(\log\log\left(\frac{1}{\epsilon}\right)\right).
\end{equation}
Additionally, if we have a circuit containing $n_{rot}$ rotations and a total error budget for gate synthesis $\epsilon_S$, we often cannot do better than bound the error required for a single rotation by the triangle inequality, such that $\epsilon \leq \frac{\epsilon_S}{n_{rot}}$. If we neglect the part in $n_T^{synth}$ that is doubly logarithmic in the inverse error and make the conservative estimates that $n_{rot} = 10^{7}$ and $\epsilon_{S} \leq 10^{-5}$\footnote{These estimates are in broad agreement with the estimates for FeMoco in~\cite{Reiher2017}. The Hamiltonian considered therein had $6.1 \times 10^6$ terms in one instance. One could consider an $\epsilon_S$ that is close to chemical accuracy, $\epsilon_S \leq 10^{-3}$~Ha, but there are other sources of error in the quantum algorithm whose T count scales less benignly and consume a larger portion of the total error budget (for example, phase estimation error).}, each of these PPRs can be synthesised from approximately 100 $T$ gates. Thus the approximate cost to implement the controlled version of $\exp\{i\gamma\tilde{H}_{JW}\}$ is roughly \textbf{1600 $T$ gates}.

\section{The gadget}
\label{sec:gadget}

Clearly, $\tilde{H}_{JW}$ isn't just a sum of arbitrary Pauli strings, and each term shares the same coefficient. The key point is to use this information to note that $\tilde{H}_{JW}$ can be rewritten as

\begin{align}
      \nonumber \tilde{H}_{JW} &= (\vert{0011}\rangle\langle{1100}\vert+\vert{1100}\rangle\langle{0011}\vert)_{pqrs} \\ &\otimes \bigotimes_{\substack{p<t<q \\ r<t<s}} Z_t \otimes \bigotimes_{\text{elsewhere}} I.
\end{align}

\begin{figure}[!h]
  \centering
  \includegraphics[height=0.65\columnwidth]{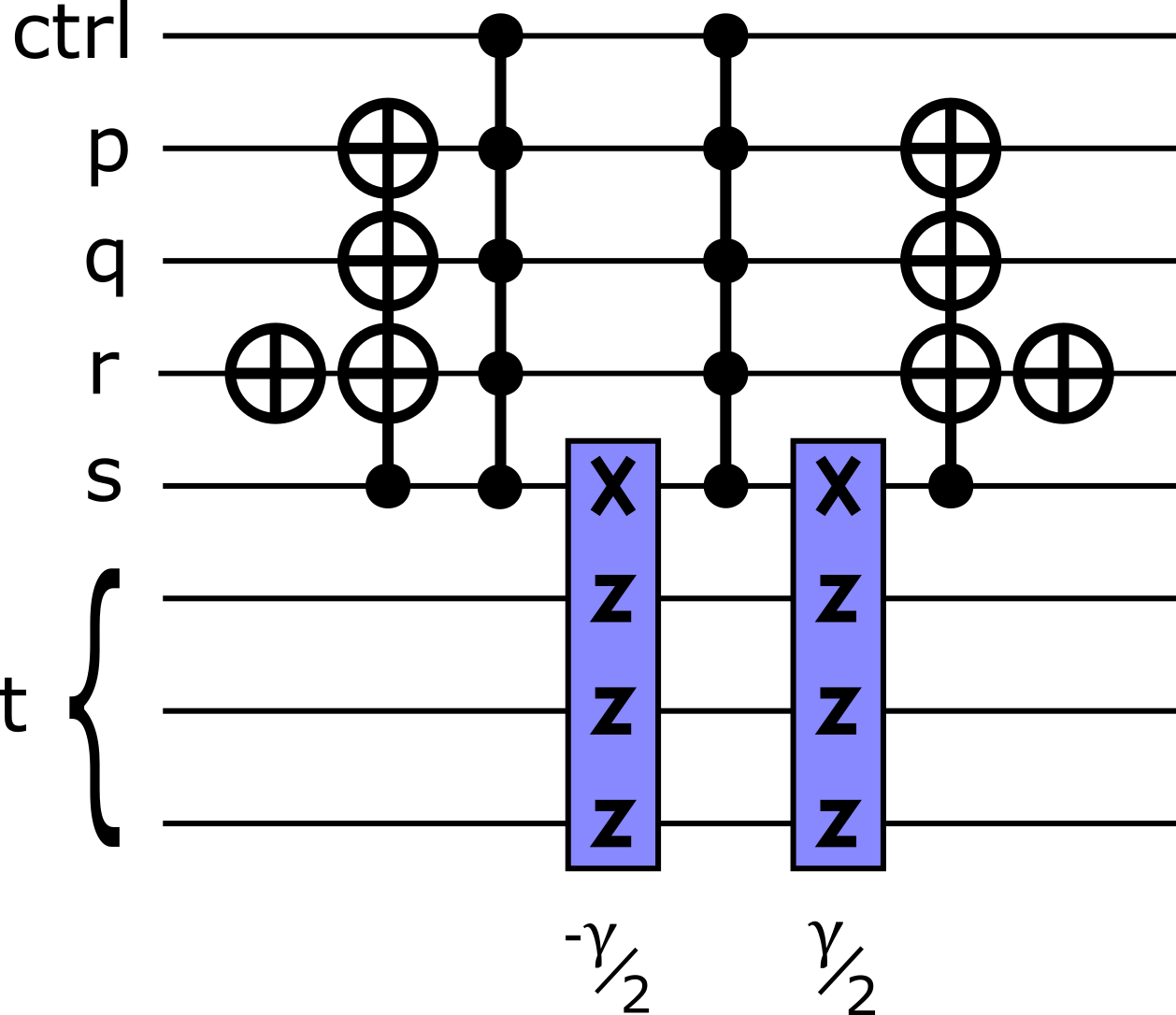}
  \caption{The gadgetized form of $\exp\{i\gamma \tilde{H}_{JW}\}$, with the controlled rotation unpacked, an additional control added, and the evolution extended to include qubits $p<t<q, r<t<s$. The PPR is drawn as a purple box containing the corresponding Pauli string.}
  \label{fig:pqrs_ppr}
\end{figure}

We give an explicit circuit for the controlled version of $\exp\{i\gamma \tilde{H}_{JW}\}$ that utilises this expression in Fig.~\ref{fig:pqrs_ppr}. However, we will first motivate its form with a few simpler cases.

\begin{figure}[!h]
  \centering
  \includegraphics[height=0.38\columnwidth]{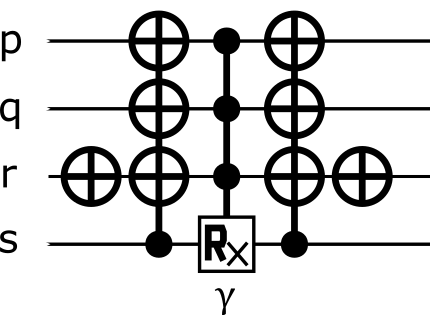}
  \caption{A circuit that acts as $\exp\{i\gamma \tilde{H}_{JW}\}$ in the subspace spanned by qubits $\{p,q,r,s\}$.}
  \label{fig:pqrs_crz}
\end{figure}

Consider first the part of the Hamiltonian that acts on the subspace spanned by qubits labelled by $\{p,q,r,s\}$. In this subspace, the unitary $\exp\{i\gamma (\vert{0011}\rangle\langle{1100}\vert+\vert{1100}\rangle\langle{0011}\vert)_{pqrs}\}$ is a multi-controlled $R_X$ rotation, up to Cliffords. The explicit circuit is shown in Fig.~\ref{fig:pqrs_crz}.

\begin{figure}[!h]
  \centering
  \includegraphics[height=0.38\columnwidth]{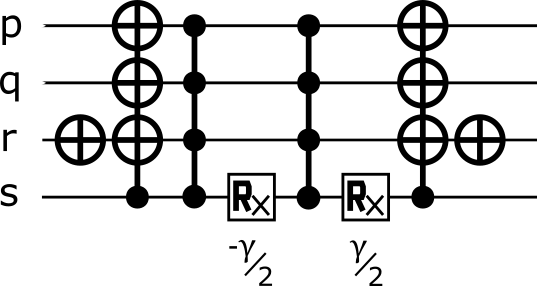}
  \caption{A circuit that acts as $\exp\{i\gamma \tilde{H}_{JW}\}$ in the subspace spanned by qubits $\{p,q,r,s\}$, with the controlled rotation unpacked.}
  \label{fig:pqrs_rz}
\end{figure}

The multi-controlled $R_X$ can be unpacked into two multi-controlled CZs (with 3 controls) and two regular $R_X$ rotations (rotating in opposite directions), as shown in Fig.~\ref{fig:pqrs_rz}.

\begin{figure}[!h]
  \centering
  \hspace{-0.6cm}
  \includegraphics[height=0.45\columnwidth]{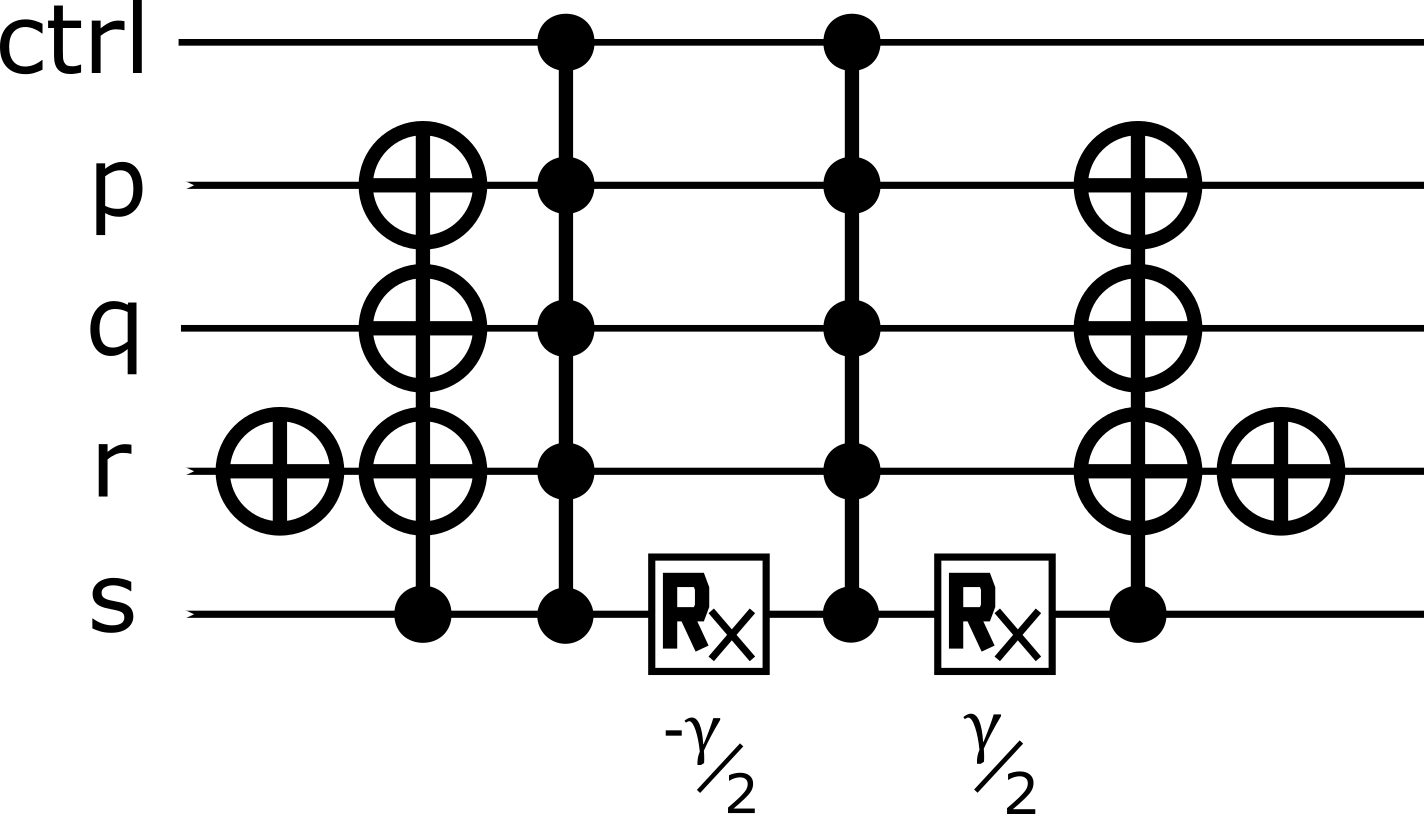}
  \caption{A circuit that acts as $\exp\{i\gamma \tilde{H}_{JW}\}$ in the subspace spanned by qubits $\{p,q,r,s\}$, with the controlled rotation unpacked and an additional control added.}
  \label{fig:pqrs_cz_ctrl}
\end{figure}

Moreover, if we're interested in a controlled version of $\exp\{i\gamma \tilde{H}_{JW}\}$, the only operations that need an additional control are the CZs, as in Fig.~\ref{fig:pqrs_cz_ctrl}.

To incorporate the missing $Z$s, we only need to replace the $R_X$ rotations with PPRs of the form $\exp\{i\gamma (XZ\ldots Z)\}$, as in Fig.~\ref{fig:pqrs_ppr}.

The total cost of this circuit in terms of PPRs is therefore 2, and synthesising to the same accuracy as previously requires 200 $T$ gates. The Toffolis also have a $T$ cost, but it is largely dwarfed by the cost of the synthesised PPRs. The cost of decomposing Toffolis depends on the number of dirty ancillae available; in this instance, we only need 2 dirty ancillae to decompose the 4-controlled Toffoli into 8 regular Toffolis, which then can be decomposed into 32 T gates using the Jones decomposition~\cite{Jones2013}. In all but the smallest instances, these ancillae will be available to exploit. Thus the total T count of the gadget is \textbf{264 T gates}, which is an approximately 6x reduction on the original approach.

\section{Summary}

We have demonstrated a compilation gadget that, given a realistic choice of parameters, reduces the cost of Trotterized quantum algorithms for quantum chemistry by 6x. Moreover, with growing problem size the advantage asymptotes to 8x. In this regime, the proportion of terms in the Hamiltonian that are of the form $a^\dagger_p a^\dagger_q a_r a_s$ tends to one, and the required number of rotations increases (and hence the accuracy per term, $\epsilon$, decreases). Hence the cost of the additional Toffolis diminishes in comparison to the T count for the increasingly precise rotations. 

\section{Acknowledgements}

The author would like to thank \mbox{Daniel} \mbox{Litinski} for initial discussion regarding this result, \mbox{Eric} \mbox{Johnston} for circuit simulation support and \mbox{Andrew} \mbox{Doherty} and \mbox{Terry} \mbox{Rudolph} for useful comments on preparing this manuscript.


\bibliographystyle{unsrt}
\bibliography{main}

\onecolumn\newpage
\appendix

\section{Gadgetizing other Hamiltonian terms}
\label{sec:other_terms}

We can gadgetise the other terms in the Hamiltonian in an identical way to the term discussed in the main body.

\subsection{$a^\dagger_p a_p$ terms}

Under Jordan-Wigner this term transforms as:
$$
a^\dagger_p a_p \rightarrow \vert 1 \rangle \langle 1 \vert_p
$$
And so its corresponding controlled unitary can be replaced with the following:

\begin{figure}[!h]
  \centering
  \vspace{0.2cm}
  \includegraphics[height=0.15\columnwidth]{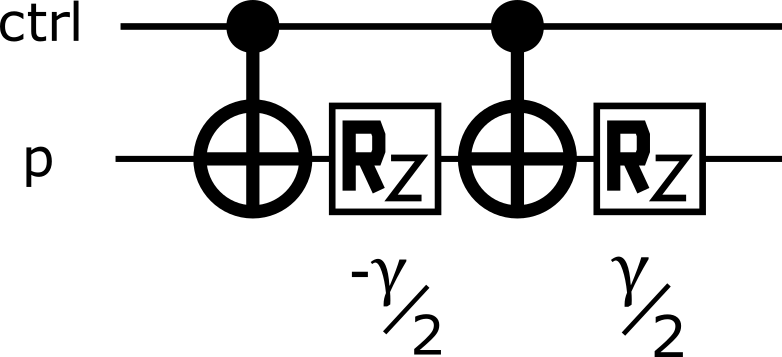}
  \caption{A gadgetized $a^\dagger_p a_p$ term.}
  \label{fig:pp}
\end{figure}

\subsection{$a^\dagger_p a_q + a^\dagger_q a_p$ terms}

Under Jordan-Wigner this term transforms as:
\begin{align}
a^\dagger_p a_q + a^\dagger_q a_p \rightarrow \frac{1}{2}(X_p \otimes X_q + Y_p \otimes Y_q) &\otimes \bigotimes_{p<t<q} Z_t \\
= (\vert 01 \rangle\langle 10\vert + \vert 10 \rangle\langle 01\vert)_{pq} &\otimes \bigotimes_{p<t<q} Z_t. 
\end{align}
And so its corresponding controlled unitary can be replaced with the following:

\begin{figure}[!h]
  \centering
  \vspace{0.2cm}
  \includegraphics[height=0.3\columnwidth]{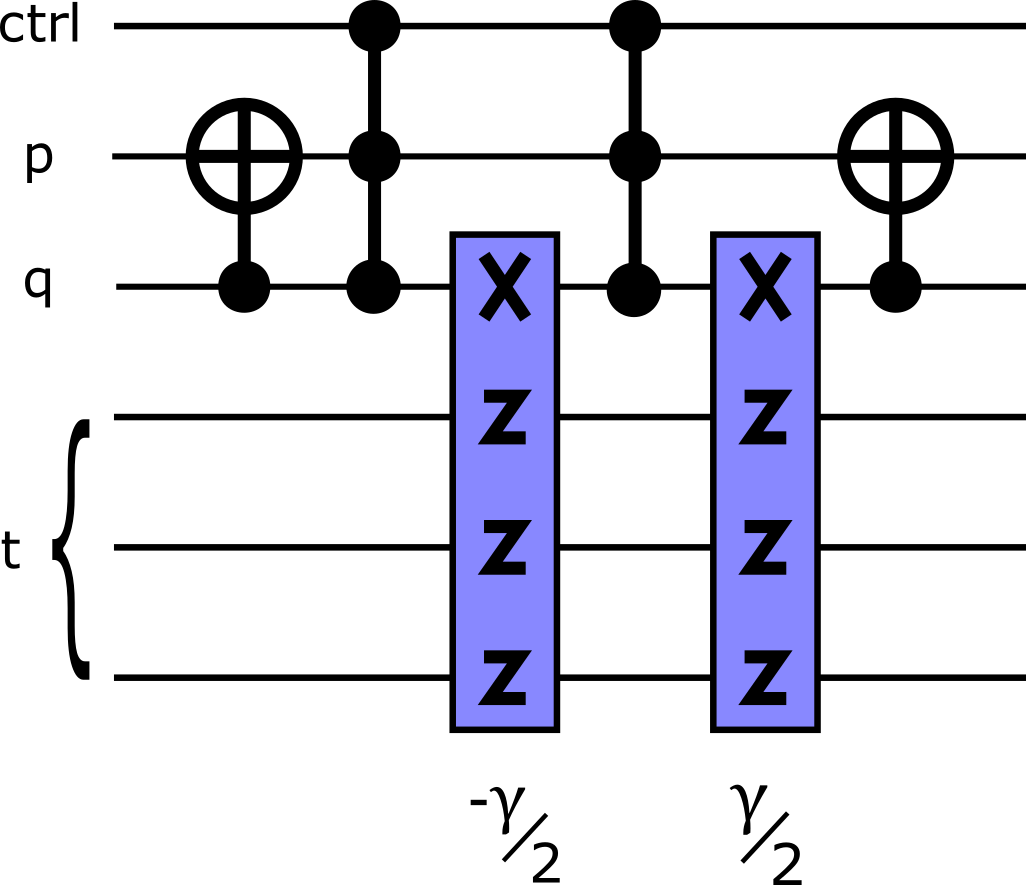}
  \caption{A gadgetized $a^\dagger_p a_q + a^\dagger_q a_p$ term.}
  \label{fig:pq}
\end{figure}

\subsection{$a^\dagger_p a^\dagger_q a_q a_p$ terms}

Under Jordan-Wigner this term transforms as:
\begin{equation}
a^\dagger_p a^\dagger_q a_q a_p \rightarrow \vert 11 \rangle\langle 11 \vert_{pq}
\end{equation}
And so its corresponding controlled unitary can be replaced with the following:

\begin{figure}[!h]
  \centering
  \vspace{0.2cm}
  \includegraphics[height=0.18\columnwidth]{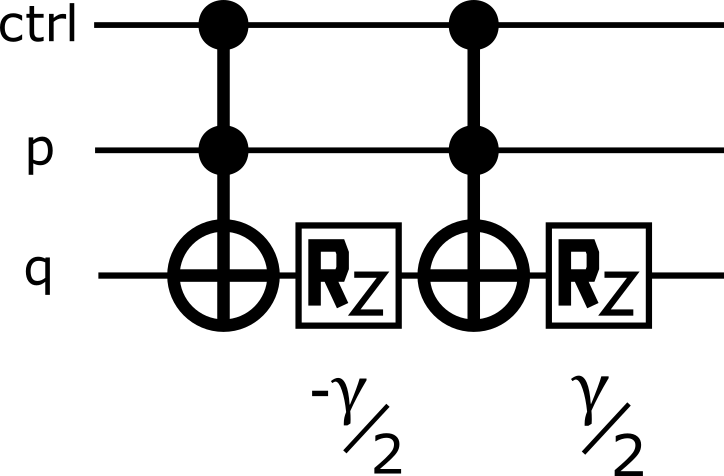}
  \caption{A gadgetized $a^\dagger_p a^\dagger_q a_q a_p$ term.}
  \label{fig:pqqp}
\end{figure}

\newpage
Note that the other two-body term with two repeated indices, $a^\dagger_p a^\dagger_q a_p a_q + a^\dagger_q a^\dagger_p a_q a_p$, is equivalent to this term up a minus sign (readily checkable by commuting the right-hand two operators), so has the same circuit decomposition but with rotations in the opposite direction.

\subsection{$a^\dagger_p a^\dagger_q a_p a_r + a^\dagger_r a^\dagger_p a_q a_p$ terms}

Assuming $q<r$ wlog, under Jordan-Wigner this term transforms as:
\begin{align}
a^\dagger_p a^\dagger_q a_p a_r + a^\dagger_r a^\dagger_p a_q a_p &\rightarrow \frac{1}{4}(X_q \otimes X_r \otimes Z_p + X_q \otimes X_r \otimes I_p + Y_q \otimes Y_r \otimes Z_p + Y_q \otimes Y_r \otimes I_p) \otimes \bigotimes_{q<t<r} Z_t \\
&= -(\vert 101 \rangle\langle 110\vert + \vert 110 \rangle\langle 101\vert)_{pqr} \otimes \bigotimes_{q<t<r} Z_t. 
\end{align}
Note that this expression is correct independent of whether $p$ is smaller or larger than either $q$ or $r$.

Therefore, its corresponding controlled unitary can be replaced with the following:

\begin{figure}[!h]
  \centering
  \vspace{0.2cm}
  \includegraphics[height=0.35\columnwidth]{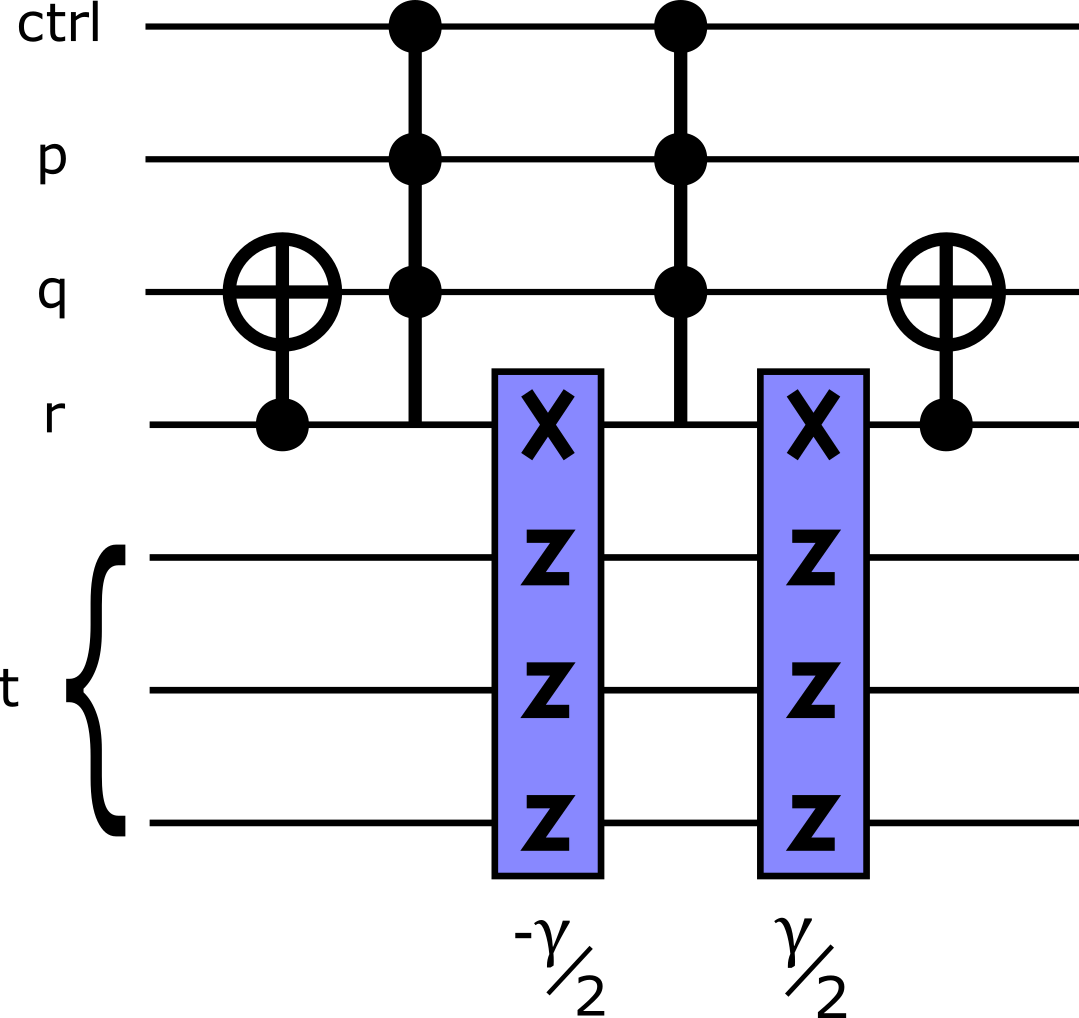}
  \caption{A gadgetized $a^\dagger_p a^\dagger_q a_p a_r$ term.}
  \label{fig:pqpr}
\end{figure}

It can be checked by exhaustive enumeration that every other possible two-body term with a single repeated index is equivalent to the term above, up to: a) commutation of creation operators; b) commutation of annihilation operators; c) relabelling of indices.

\end{document}